\documentclass[preprint]{emulateapj}
\usepackage{epsf}
\usepackage{apjfonts}
\usepackage{xspace}
\usepackage{amsmath}
\begin{document}

\shorttitle{Virial Scaling Relations of Groups and Clusters}
\submitted{The Astrophysical Journal, submitted}
\slugcomment{{\em The Astrophysical Journal, submitted}} 
\shortauthors{LAU, NAGAI \& KRAVTSOV}

\title{Effects of Baryon Dissipation on the Dark Matter Virial Scaling Relation}

\author{Erwin T. Lau\altaffilmark{1}, Daisuke Nagai\altaffilmark{2,3}, Andrey V. Kravtsov\altaffilmark{1,4}}

\altaffiltext{1}{Department of Astronomy and Astrophysics, 5640 South Ellis Ave., The
  University of Chicago, Chicago, IL 60637 ({\tt ethlau@oddjob.uchicago.edu})} 
\altaffiltext{2}{Department of Physics, Yale University, New Haven, CT 06520}
\altaffiltext{3}{Yale Center for Astronomy \& Astrophysics, Yale University, New Haven, CT 06520}
\altaffiltext{4}{Kavli Institute for Cosmological Physics and
  Enrico Fermi Institute, 5640 South Ellis Ave., The University of
  Chicago, Chicago, IL 60637}

\begin{abstract}
We investigate effects of baryon dissipation on the dark matter virial
scaling relation between total mass and velocity dispersion and the
velocity bias of galaxies in groups and clusters using self-consistent
cosmological simulations.  We show that the baryon dissipation
increases the velocity dispersion of dark matter within the virial
radius by $\approx 5-10\%$.  The effect is mainly driven by the change
in density and gravitational potential in inner regions of cluster,
and it is larger in lower mass systems where gas cooling and star
formation are more efficient.  We also show that the galaxy velocity
bias depends on how galaxies are selected.  Galaxies selected based on
their stellar mass exhibit no velocity bias, while galaxies selected
based on their total mass show positive bias of $\approx 10\%$,
consistent with previous results based on collisionless dark matter-only 
simulations.  We further find that observational estimates of
galaxy velocity dispersion are unbiased with respect to the velocity
dispersion of dark matter, provided galaxies are selected using their
stellar masses and and their velocity dispersions are computed with more
than twenty most massive galaxies.  Velocity dispersions estimated with
fewer galaxies, on the other hand, can lead to significant
underestimate of dynamical masses.  Results presented in this paper
should be useful in interpretating high-redshift groups and clusters
as well as cosmological constraints derived from upcoming optical
cluster surveys.
\end{abstract}
 
\section{Introduction}
\label{sec:intro}

Galaxy clusters are unique laboratories for astrophysics and powerful
probes in cosmology.  As the largest and most massive
gravitationally-bound structures in the universe, they provide
well-defined environments for studies of various astrophysical
processes: e.g., plasma interactions in the intracluster medium,
galaxy interactions and evolution, supermassive black hole feedback
processes, etc. In the cosmological context, galaxy clusters are good
tracers of the cosmological evolution, as their overall dynamics are
largely driven by gravity \citep[][see \citeauthor{voit05}
  \citeyear{voit05} for a review]{kaiser86}. Different ways in which
clusters can be used for cosmology rely on understanding of how their
observable properties are related to the total virialized mass.  One
of the main challenges of cluster cosmology and astrophysics are thus
understanding this relation and accurate measurements of cluster
mass in observations.

Dynamical mass estimate uses motions of cluster galaxies as tracers of
the mass distribution of groups and clusters, and it is a relatively
inexpensive and efficient technique.  The
first dynamical measurement of cluster mass dates back to the
pioneering works of \citet{zwicky33,zwicky37} and \citet[][see
  \citeauthor{biviano00} \citeyear{biviano00} for a historical
  overview]{smith36} .  To date, a variety of dynamical mass
estimation techniques has been devised and used extensively,
including the projected mass estimator
\citep{heisler_etal85,rines_etal03,rines_diaferio_06}, the Jeans
analysis
\citep{carlberg_etal97,girardi_etal98,van_der_Marel_etal00,biviano_girardi_03,rines_etal03,katgert_etal04,
  biviano_poggianti_09}, and the caustic method
\citep{diaferio_geller_97,diaferio_99,biviano_girardi_03,rines_etal03,
  rines_diaferio_06,rines_etal07,diaferio_etal05}. For low-mass
clusters and groups ($M\lesssim 10^{14}\,\rm M_{\odot}$), especially
at higher redshifts, dynamical mass estimate is often the only 
available approach to obtain mass \citep[e.g.,][]{coil_etal06}, 
since they are mostly X-ray dim and not massive enough to be detected by the
Sunyaev-Zeldovich effect.  Dynamical mass estimates also provide
important check on mass measurements obtained using other methods and
help in calibration of optical mass-richness relation
\citep[e.g.][]{becker_etal07}.

Recently, \citet{evrard_etal08} used a large set of dissipationless
$N$-body cosmological simulations (some with non-radiative
gasdynamics) and showed that virial scaling relation for dark matter
particles is consistent with predictions of the virial theorem, in
which the velocity dispersion scales with halo mass to the power of
$1/3$.  Their study further showed that the virial scaling relation
exhibits remarkable degree of universality and self-similarity across
a broad range of halo masses, redshifts, and cosmological models, as
well as having very small scatter.  This is good news for the use of
optical clusters as sensitive cosmological probes.

In practice, however, applying the virial scaling relation for
precision cosmology faces a number of difficulties. First, unbiased
selection of cluster member galaxies (i.e., removal of foreground and
background interlopers) is difficult and can be a major source of
systematic uncertainty in velocity dispersion mesaurement of
cluster members \citep[see
  e.g.,][]{mckay_etal02,biviano_etal06,chen_etal06}, especially for
low-mass clusters and groups consisting of only a few bright
galaxies \citep[e.g.,][]{hernquist_etal95}.  Second, since we can only
measure the line-of-sight velocity, velocity anisotropy of
galaxies is also an additional source of systematic uncertainty
\citep[e.g.,][]{biviano_poggianti_09}.  In order to use velocity
dispersion as an accurate and precise tracer of cluster mass for future
optical cluster surveys, it is imperative to understand and control
these uncertanties.

However, even if cluster member selection and velocity
anisotropies are well understood and controlled, we need to understand
possible systematic uncertainties in the intrinsic relation between
three-dimensional velocity dispersion of galaxies and total cluster
mass. For example, the simulations used in the study of
\citet{evrard_etal08} did not include effects of dissipative baryonic
processes, which do accompany cluster formation and lead to formation
of cluster galaxies. From a number of X-ray observations we know that
self-similarity is broken in cluster cores and group-size systems and
this causes deviations of scaling relations from theoretical
predictions of cluster formation where gravity is the only energy
driver
\citep[e.g.,][]{markevitch_98,arnaud_evrard_99,osmand_ponman_04,pratt_etal09}.
Baryon dissipation, in particular, modifies the total mass profile
and, likely, the velocity dispersion profile in groups and clusters
both due to the contraction and deepening of cluster potential
\citep[e.g.,][]{zeldovich_etal80,blumenthal_etal86,GKKN04,sellwood_mcgaugh05}
and due to the overall redistribution of baryonic mass
\citep[e.g.,][]{rudd_etal08}.  Observationally, we can only measure
velocity dispersions of cluster galaxies, which can be a biased tracer
of dark matter velocity dispersion.  This {\it velocity bias\/} is
also a potential source of systematic uncertainties
\citep[e.g.,][]{carlberg90,evrard_etal94,frenk_etal96,col_etal00,ghigna_etal00,
  springel_etal01,diemand_etal04,gao_etal04,faltenbacher_etal05,biviano_etal06,faltenbacher_etal06}.

In this paper, we focus on the latter two intrinsic systematic uncertainties. 
Namely, we investigate the effect of dissipation on the dark matter virial scaling relation and
velocity bias of cluster galaxies using high resolution self-consistent cosmological cluster simulations. 
We show that the baryon dissipation increases the velocity dispersion of dark matter, by changing the inner 
density profiles and potential of gravitationally bound systems, and that the effects are larger 
in lower mass systems where gas cooling and star formation are more efficient.  We also show that 
observational estimate of dynamical mass gives unbiased mass estimate of galaxy groups and clusters, 
in contrary to the previous results obtained based on dark matter only simulations.  We describe our
simulations in \S~\ref{sec:sim} and report results in \S~\ref{sec:results}.  Main results are summarised
in \S~\ref{sec:summary}.

\section{The Simulation}
\label{sec:sim}

In this study, we analyze high-resolution cosmological simulations of 16 cluster-sized 
systems performed using the Adaptive Refinement Tree (ART) $N$-body$+$gasdynamics 
code \citep{kra99,kra02}. The ART is a Eulerian code that uses adaptive refinement in space 
and time, and (non-adaptive) refinement in mass \citep{klypin_etal01} to reach the high dynamic
range required to resolve cores of halos formed in self-consistent cosmological simulations. 
The simulations presented here are discussed in detail in \citet{nag07b} and \citet{nag07a} 
and we refer the reader to these papers for more details.  Here we highlight aspects of the 
simulations relevant for this work.  

\begin{table}[t]
\caption{Properties of the simulated CSF clusters at $z=0$}
\begin{center}
\begin{tabular}{lccc}
\hline
\hline
Cluster ID \hspace*{5mm}&$r_{200c}$&$M_{200c}$&$N(<r_{200c})$\\
&[$h^{-1}$ Mpc]&[$10^{14}\,h^{-1}$M$_{\odot}$]\\
\hline
CL101 \dotfill &1.786 & 13.2 & 1243854\\
CL102 \dotfill &1.509 & 7.98 & 749369\\
CL103 \dotfill &1.585 & 9.24 & 867435\\
CL104 \dotfill &1.435 & 6.87 & 633240\\
CL105 \dotfill &1.443 & 7.00 & 649012\\
CL106 \dotfill &1.316 & 5.30 & 496021\\
CL107 \dotfill &1.174 & 3.76 & 348465\\
CL3 \dotfill &1.104 & 3.13 & 981098\\
CL5 \dotfill &0.925 & 1.84 & 574968\\
CL6 \dotfill &0.958 & 2.05 & 644496\\
CL7 \dotfill &0.937 & 1.91 & 603447\\
CL9 \dotfill &0.657 & 0.66 & 204249\\
CL10 \dotfill &0.710 & 0.83 & 261205\\
CL11 \dotfill &0.779 & 1.09 & 338860\\
CL14 \dotfill &0.755 & 1.00 & 312626\\
CL24 \dotfill &0.671 & 0.70 & 221012\\
\hline
\end{tabular}
\end{center}\label{tab:cldat}
\end{table}

The $N$-body$+$gasdynamics cluster simulations used in this analysis include collisionless 
dynamics of dark matter and stars, and gasdynamics.  In order to assess the effects of gas cooling 
and star formation on the dynamics of dark matter, we conducted each cluster simulation with
two different prescriptions for gasdynamics: one with only the standard gasdynamics for the 
baryonic component without radiative cooling and star formation - the `non-radiative' (NR) runs, 
and  `cooling+SF' (CSF) runs.  In the CSF runs, several physical processes critical to various 
aspects of galaxy formation are included: star formation, metal enrichment and thermal feedback 
due to supernovae Type II and Type Ia, self-consistent advection of metals, metallicity dependent 
radiative cooling and UV heating due to cosmological ionizing background (see  \citet{nag07b} 
for details of the metallicity-dependent radiative cooling and star formation).  These simulations 
therefore follow the formation of galaxy clusters starting from the well-defined cosmological initial 
conditions and capture the dynamics and properties of the ICM in a realistic cosmological context.  
However, some potentially relevant physical processes, such as AGN bubbles, magnetic field, and 
cosmic rays, are not included.  Therefore, the simulated cluster properties are probably not fully 
realistic in the innermost cluster regions, where these processes are likely important. 

Our simulated sample includes 16 clusters at $z=0$ and their most massive progenitors at $z=0.6$ 
and $z=1.0$. The properties of simulated clusters at $z=0$ are given in Table~\ref{tab:cldat}. 
The total cluster masses are reported at the radius $r_{200c}$ enclosing overdensities with respect 
to the critical density at the redshift of the output (below, we also use a higher overdensity level, 500). 
The dark matter (DM) particle mass in the region around the cluster was $9\times 10^{8}\,h^{-1}\, 
{\rm M_{\odot}}$ for CL101-107 and $3\times 10^{8}\,h^{-1}\,{\rm M_{\odot}}$ for CL3-24, while other 
regions were simulated with lower mass resolution.  The peak resolution of the cluster simulation 
is $\approx 2-3 \,h^{-1}$~kpc in the box size of 80-120$\,h^{-1}$~Mpc.   The simulations are performed
for the flat {$\Lambda$}CDM model: $\Omega_{\rm m}=1-\Omega_{\Lambda}=0.3$, 
$\Omega_{\rm b}=0.04286$, $h=0.7$ and $\sigma_8=0.9$, where the Hubble constant is defined 
as $100h{\ \rm km\ s^{-1}\ Mpc^{-1}}$, and an $\sigma_8$ is the power spectrum normalization 
on an $8h^{-1}$~Mpc scale.

\section{Results}
\label{sec:results}
\subsection{Virial Scaling Relation}
\label{sec:scaling}

\begin{table}[t]
\caption{Scaling Relation Parameters}
\begin{center}
\begin{tabular}{l|cccc}
\hline
\hline
&&&$\Delta_c = 200$&\\
\cline{2-5}
&&$z=0.0$&$z=0.6$&$z=1.0$\\
\hline
$\alpha$&NR &$0.3072\pm0.0114$&$0.3214\pm0.0246$&$0.3512\pm0.0152$\\
        &CSF&$0.2724\pm0.0149$&$0.2903\pm0.0254$&$0.3232\pm0.0153$\\
\hline
$\sigma_{\mathrm{DM},0}$&NR &$643\pm7$&$648\pm11$&$645\pm8$\\
$\mathrm{[km/s]}$&CSF&$692\pm11$&$674\pm11$&$671\pm8$\\
\hline
$\left<\delta_{\ln\sigma}^2\right>^{1/2}$
&NR &$0.0410\pm0.0079$&$0.0630\pm0.0086$&$0.0348\pm0.0062$\\
&CSF&$0.0557\pm0.0092$&$0.0569\pm0.0098$&$0.0375\pm0.0080$\\
\hline
\hline
&&&$\Delta_c = 500$&\\
\cline{2-5}
&&$z=0.0$&$z=0.6$&$z=1.0$\\
\hline
$\alpha$&NR &$0.3335\pm0.0096$&$0.3256\pm0.0207$&$0.3475\pm0.0110$\\
        &CSF&$0.2965\pm0.0144$&$0.2888\pm0.0180$&$0.3134\pm0.0108$\\
\hline
$\sigma_{\mathrm{DM},0}$&NR &$742\pm6$&$737\pm10$&$751\pm10$\\
$\mathrm{[km/s]}$&CSF&$788\pm10$&$762\pm9$&$771\pm11$\\
\hline
$\left<\delta_{\ln\sigma}^2\right>^{1/2}$
&NR &$0.0310\pm0.0059$&$0.0444\pm0.0082$&$0.0258\pm0.0062$\\
&CSF&$0.0459\pm0.0075$&$0.0394\pm0.0073$&$0.0361\pm0.0070$\\
\hline
\end{tabular}
\end{center}\label{tab:scaling_param}
\end{table}

\begin{figure*}[t]
\epsscale{1.1}
\plotone{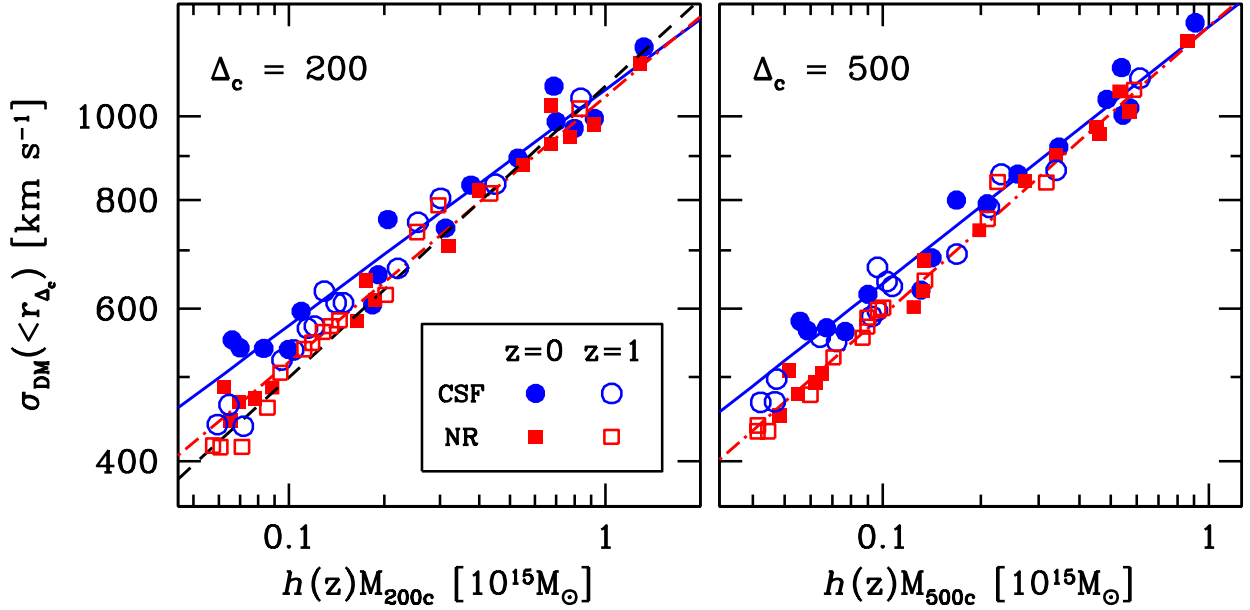}
\caption{{\em Left} panel shows the dark matter virial scaling relation for the sixteen clusters 
for $\Delta_{c} = 200$ using least-square fitting at $z=0$ and $z=1$ and the {\em right} panel shows
the same relation for $\Delta_{c} = 500$. For each panel, the CSF
clusters at $z=0$ are represented by the solid blue circles, and the CSF clusters at $z=1$ are
represented by open blue circles. The solid blue line is the best least-square fit for the 
$z=0$.  For the NR run, the $z=0$ and $z=1$ clusters
are represented by solid red squares and open red squares respectively, which the red dot-dashed line is 
the best least-square fit to the clusters at $z=0$. 
For reference, the best fit from \citet{evrard_etal08} is shown as black dashed line. 
Please refer to Table~\ref{tab:scaling_param} for the values of the fitted parameters. 
\label{fig:scaling}}
\end{figure*}

Following \citet{evrard_etal08}, we assume velocity isotropy and calculate the one-dimensional 
dark matter velocity dispersion $\sigma_{\mathrm{DM}}$ to be
\begin{equation}
\sigma_{\mathrm{DM}}^2 = \frac{1}{3N}\sum_{i=1}^3\sum_{p=1}^N\left(v_{p,i}-\bar{v}_i\right)^2,
\label{eq:sigma_DM}
\end{equation}
where $N$ is the total number of dark matter particles inside $r_{\Delta_c}$, the radius which encloses matter having 
average density $\Delta_c$ times the critical density; $v_{p,i}$ is the the $i$-th velocity component ($i=x,y,z$) 
of the $p$-th dark matter particle inside $r_{\Delta_c}$; and $\bar{v}_i$ is the peculiar velocity of the cluster is defined to be the mass-weighted average of all dark matter particles within $r_{\Delta_c}$.  We have also calculated the cluster velocity using velocities of all species (dark matter, stars and gas), and it agrees with the average velocities of dark matter to within a few percent for most clusters, since baryons are only a minor fraction of the total cluster mass. We weigh the
mean and variance by dark matter particle mass but it is insensitive to whether mass-weighted or number-weighted 
is used within $r_{\Delta_c}$. We take the cluster center to be the location of the subhalo which has the 
highest binding energy. 

A power law is then fitted to the dark matter virial scaling relation using standard least square minimization:
\begin{equation}
\log\sigma_{\mathrm{DM}}(M,z) = \log\sigma_{\mathrm{DM},0}+\alpha\log\left(\frac{h(z)M_{\Delta_c}}{2\times10^{14}{\mathrm M}_{\odot}}\right)
\label{eq:scaling}
\end{equation}
where $h(z) = H(z)$/100 km s$^{-1}$ Mpc$^{-1}$ is the normalized Hubble parameter which takes into 
the account of the redshift dependence of $M_{\Delta_c}$.  Figure~\ref{fig:scaling} shows the fitted velocity dispersion-mass relations 
of the CSF and the NR runs at $z=0$, $0.6$, and $1.0$.  Table~\ref{tab:scaling_param} 
reports the best fit parameters of the velocity dispersion-mass relations measured at $\Delta_{c}=200$ and $500$.  
The errors are estimated by generating $5000$ bootstrap samples from the sixteen clusters at each redshift.  

At $\Delta_{c}=500$, the NR scaling relation is consistent with the self-similar scaling relation 
with no evolution. At $\Delta_{c}=200$, the best-fitted NR scaling relation at $z=0$ differs slightly ($2.6\sigma$) 
from that of \citet{evrard_etal08} with $\alpha=0.3361\pm0.0026$ and 
$\sigma_{\mathrm{DM}}(M_{200c}=10^{15}\,h^{-1}\mathrm{M}_{\odot}) = 1082.9\pm 4.0$ km/s.
Note that results are fully consistent with the \citet{evrard_etal08} relation if we limit our analyses 
to systems with the mass range ($M > 10^{14}$ $h^{-1}M_{\odot}$) adopted in their analysis.  Larger
number of simulated clusters is needed to explore remaining minute deviation from the self-similar relation.

\begin{figure}[t]
\epsscale{1.2}
\plotone{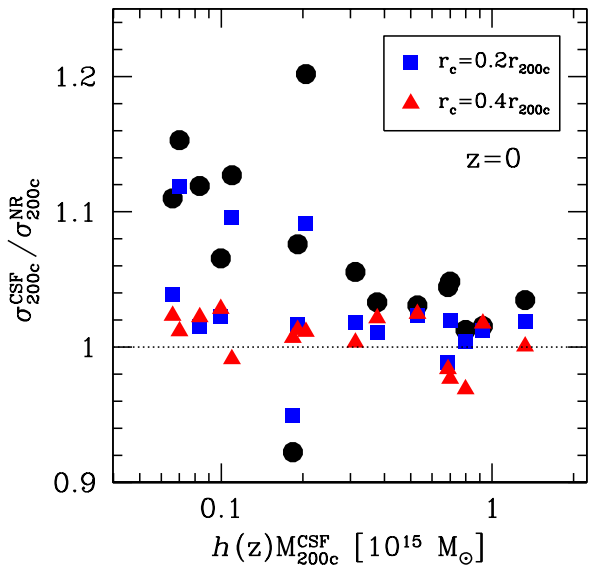}
\caption{The ratios of velocity dispersions in the CSF and NR runs, 
$\sigma^{\rm CSF}_{200c}/\sigma^{\rm NR}_{200c}$ as a function of cluster mass at $z=0$.  
Different symbols indicate velocity dispersions computed using DM particles in different 
annulus: [0-1] ({\it circles}), [0.2-1] ({\it square}) and [0.4-1] ({\it triangles}) in unit of $r_{200c}$.
\label{fig:mass_sigma_rc0}}
\end{figure}

The CSF scaling relations differ significantly from the NR relation at all redshifts. 
The baryon dissipation increases the velocity dispersion of dark matter, making the slope 
shallower in the CSF run compared to that of the NR run.  Interestingly, the log-normal scatter is only 
3-5\%, and it is remarkably insensitive to input gas physics and redshift. 

The breaking of self-similarity by baryon dissipation in the CSF runs
is illustrated more clearly in Figure~\ref{fig:mass_sigma_rc0}, which
shows the ratio of dark matter velocity dispersion in the CSF run
relative to the NR run,
$\sigma^{\mathrm{CSF}}_{200c}/\sigma^{\mathrm{NR}}_{200c}$~\footnote{As
the same cluster in CSF and NR runs may have slightly different
$M_{200c}$, we normalize each $\sigma_{200c}$ with the circular
velocity $v_{200c}\equiv \sqrt{GM_{200c}/r_{200c}}$. } as a function
of halo mass at $z=0$ on a cluster-by-cluster basis.  For almost all
halos, the inclusion of baryonic physics leads to an increase in
dark matter velocity dispersion, and the effect is greater for lower
mass systems.  We find that this bias is of the order $\sim 10\% -
15\%$ for $M_{200c}\simeq 6\times 10^{13} h^{-1}\mathrm{M}_{\odot}$,
compared to $6\%$ for $M_{200c}>3 \times 10^{14}
h^{-1}\mathrm{M}_{\odot}$.  This is because gas in lower mass halos
is at a lower virial temperature and thus capable of cooling more
efficiently to form stars.  We find the same trend at higher redshift.
Note that there is one outlier (CL5) which shows a decrease in the
velocity dispersion in the CSF run. This is due to fast moving
subhalos in the NR run that have already merged with the central
subhalo in the CSF run.  Another outlier in the opposite direction
(CL6) has a $\sim20\%$ increase in velocity dispersion. This is also 
due to a similar merging event in the CSF, boosting its velocity dispersion.

We further show that the change in velocity dispersion is largely due to the baryon dissipation 
in the central regions of clusters.  To illustrate this, we also compute velocity dispersions 
by excluding the central regions of clusters in Figure~\ref{fig:mass_sigma_rc0}.   This shows that
differences in velocity dispersions in the CSF and NR runs become smaller when we apply central exclusion.  
Excluding $r<0.2 r_{200}$ reduces the differences in the CSF and NR to about 4\% on most systems.
The bias is even smaller (less than 2\%) on all systems if we exclude the central $r<0.4 r_{200}$.  
This illustrates that the increase in $\sigma_{200c}$ is mainly driven by the increase in velocities 
in the inner regions of CSF groups and clusters. 

\begin{figure}[t]
\epsscale{1.2}
\plotone{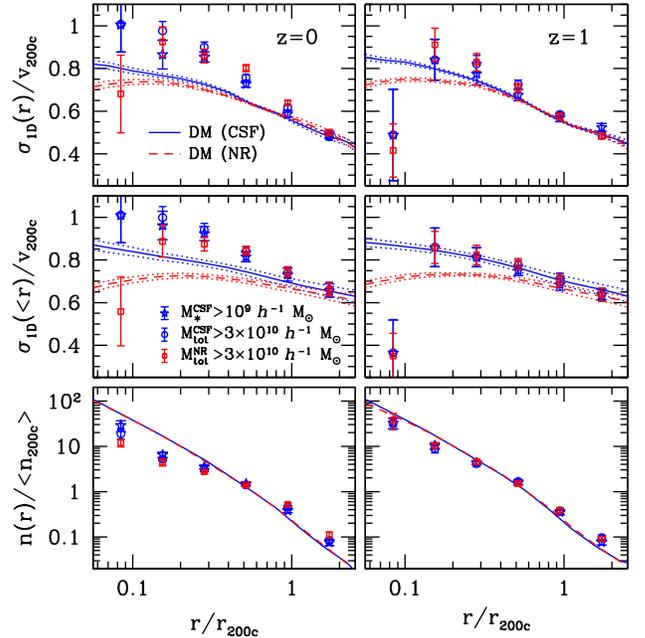}
\caption{Average DM velocity dispersion profiles for the sixteen clusters at $z=0$ ({\em left-panels})
and $z=1$ ({\em right-panels}) for CSF (blue solid lines) and NR (red dashed lines) runs. 
Dotted lines are the corresponding 1$\sigma$ errors. 
The {\em top} panels show 1-D differential velocity dispersion profiles; 
the {\em middle} panels show the 1-D cumulative velocity dispersion profiles; 
the {\em bottom} panels show the corresponding number density profiles. 
Also plotted are profiles for CSF cluster galaxies (blue star), 
CSF subhalos (blue circle) and NR subhalos (red squares).\label{fig:velpro}}
\end{figure}

Figure~\ref{fig:velpro} shows the velocity dispersion profiles averaged over the 16 clusters at $z=0$
and $z=1$ for both CSF and NR runs. Note that the mass depedence seen in Figure~\ref{fig:mass_sigma_rc0}
is explicity taken out by dividing the circular velocity $v_{200c}$.  The dark matter velocity dispersion 
of CSF clusters increases towards smaller radii compared to their NR counterparts. The change in the shape of 
the velocity dispersion profile is a direct consequence of the enhancement of dark matter density caused 
by adiabatic contraction reported previously based on our simulations \citep{GKKN04}: as gas cools and 
condenses towards the center, it deepens the central potential and drags dark matter inward. 

Note that we have assumed dark matter velocity disperion to be
isotropic. This, however, is not a completely valid assumption.  In fact, some
level of velocity anisotropy is expected for collisionless dark matter
particles
\citep{cole_lacey_96,carlberg_etal97,col_etal00,hansen_moore_06}.
Cluster galaxies are also collisionless as they have relaxation time
greater than the Hubble time.  Velocity anisotropy is therefore an
important ingredient in the dynamical mass estimates. The anisotropy
is commonly quantified by the anisotropy parameter $\beta$, defined as
\begin{equation}
\beta\equiv 1-0.5\sigma_t^2/\sigma_r^2,
\end{equation}
where $\sigma_r$ and $\sigma_t$ are the radial and tangential velocity 
dispersions respectively. The definition of $\sigma$ follows from Eq.~(\ref{eq:sigma_DM}),
where contribution of mean velocities, which is non-zero beyond the virial radius, is included.  

\begin{figure}[t]
\epsscale{1.2}
\plotone{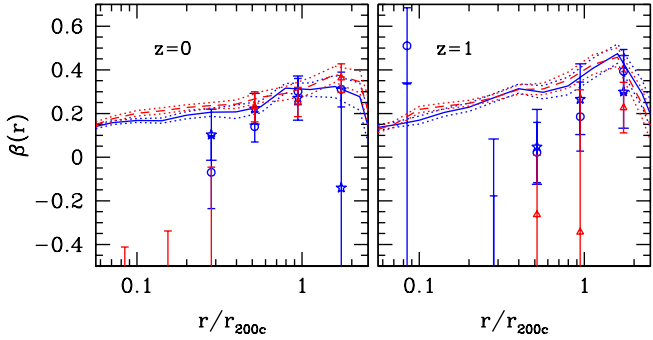}
\caption{The average velocity anisotropy profiles. The notations are the same as Figure~\ref{fig:velpro}.
\label{fig:velani}}
\end{figure}

We therefore use our simulations to study the velocity anisotropy of dark 
matter as well as galaxies in groups and clusters.  Figure~\ref{fig:velani} 
shows that dark matter velocity is nearly isotropic in the inner region of 
halo at $0.1r_{200c}$, and it becomes increasingly radial at larger 
radii 
(e.g., $\beta=0.3$ at $0.5r_{200c}$ and $\beta=0.35$ at $r_{200c}$). 
Interestingly, we find that the dark matter velocity anisotropy profile is almost
unaffected by the addition of cooling, star formation and feedback and insensitive 
to redshift between $z=0$ and $1$.  The CSF velocity anisotropy is only slightly lower
than that of the NR one. This makes sense given that the velocity anisotropy 
profile is only dependent on the slope of the density profile at large radii \citep{hansen_moore_06}, 
where the effect of baryon dissipation is small.

Finally, \citet{evrard_etal08} show that low resolution can bias the
slope of the scaling relation high. Their resolution studies indicate
that the calibration of the virial scaling relation is not affected by
resolution if clusters are resolved with more than $10^5$ dark matter
particles.  In this work, we ensured that all groups and clusters
used in our analyses are resolved with greater than $3\times 10^5$
particles.  Thus, the resolution of our simulations is sufficient for
the virial scaling relation work.  The magnitude of the baryon
dissipation, on the other hand, depends sensitively on the cooled gas
fractions in simulations.  Given that the current simulations suffer
from the overcooling problem \citep[e.g.,][]{borgani_kravtsov09}, the
effect of baryon dissipation in this work should be taken as an {\it
upper} limit.

\subsection{Velocity Bias}
\label{sec:vbias}

Since dark matter velocities are not directly observable, galaxies are used as tracers of 
the dynamics of underlying dark matter.  Velocity dispersion of cluster galaxy $\sigma_{\mathrm{gal}}$, 
however, is believed to be a ``biased'' tracer of the dark matter dynamics, $\sigma_{\mathrm{DM}}$ 
\citep{carlberg90,evrard_etal94,frenk_etal96,col_etal00,ghigna_etal00,springel_etal01,faltenbacher_etal05,faltenbacher_etal06,benatov_etal06}.
This velocity bias is often defined as,
\begin{equation}
b_v \equiv \frac{\sigma_{\mathrm{gal}}}{\sigma_{\mathrm{DM}}},
\label{eq:bv}
\end{equation}
where, as before, the one-dimensional velocity dispersions of both galaxies
and dark matter are calculated assuming isotropy (i.e., Eq.~(\ref{eq:sigma_DM})).

\begin{figure}[t]
\epsscale{1.2}
\plotone{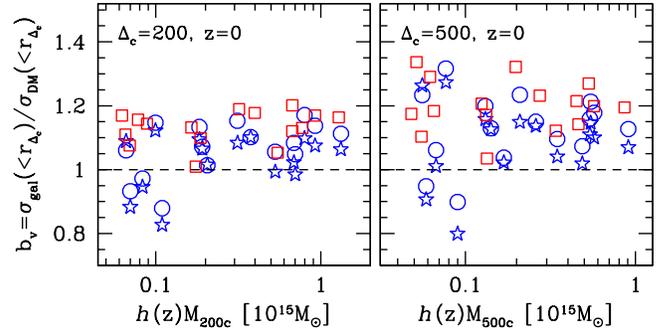}
\caption{ {\em Left} panel shows the velocity bias for $\Delta_{c} = 200$ at $z=0$, while the
{\it right} panel shows the same for $\Delta_{c} = 500$. 
For each panel, the red squares represent the velocity bias for the NR subhalos selected using their
current  total bound mass of
$\geq 3 \times 10^{10}\,h^{-1}\mathrm{M}_{\odot}$, the blue circles are velocity bias for CSF subhalos selected with 
the same mass threshold. Blue stars show the velocity bias for the CSF subhalos
selected using their stellar masses with threshold $M_{\ast}\geq 10^9\,h^{-1}\mathrm{M}_{\odot}$. 
We refer reader to the Table~\ref{tab:vbias} for the mean velocity bias and $1-\sigma$ scatter at higher
redshifts.
\label{fig:galbias}}
\end{figure}

Figure~\ref{fig:galbias} shows the velocity bias for both the CSF and NR
runs.  Following \citet{nag05}, we select subhalos using
their total bound mass ($M_{\mathrm{tot}} \geq 3 \times
10^{10}\,h^{-1}\mathrm{M}_{\odot}$) and stellar mass ($M_\ast \geq
10^9\,h^{-1}\mathrm{M}_{\odot}$).  In the NR runs, we only have the first
sample, since there are no stars in these simulations.  Results are
summarized in the Table~\ref{tab:vbias}.

A main finding is that the velocity bias depends on how
galaxies are selected.  In the NR runs, we select subhalo
using their current total bound mass and find that there is a
positive bias of about $10\%$ at all redshifts.  This is in good agreement
with the positive velocity bias of $\approx 10\% - 15\%$ for dark matter
subhalos in $N$-body simulations \citep{col_etal00,diemand_etal04,gao_etal04}.  
This positive velocity bias
arises because selection of subhalos using their current bound mass radially biases 
the resulting sample to larger radii \citep{nag05}. This selection also biases 
the sample by including more recently accreted subhalos which have larger velocity 
dispersions.  Tidal disruption can lead to not only mass loss but also complete 
disruption (or, equivalently, to mass loss that brings subhalo mass below 
resolution of simulation). The disrupted subhalos tend to be biased towards
smaller radii and cannot be recovered by any selection criterion. It is generally
thought that dissipation should enhance the survival of subhalos, although recent simulations
show that effect is not as large as has been previously thought \citep{nag05,maccio_etal06}.  
Indeed, the numbers in Table~\ref{tab:vbias} show that using
the same subhalo selection criteria in the CSF runs the 
velocity bias is noticeably smaller (by a few percents).  This means that
baryon dissipation makes subhalos somewhat less susceptible to
tidal disruption in the inner regions of groups and clusters.  

In the CSF runs, we also select galaxies based on their stellar mass.  
Interestingly, galaxies selected this way shows almost {\it no} velocity bias.  
The selection criteria are important here because stars formed in the inner regions of galactic-size
halos are tightly bound to the system and are more resistant to tidal mass
loss \citep{nag05}.  This also explains why the bias
is generally smaller at higher redshifts, where galaxies has not yet
had the time to experience significant tidal mass loss \citep[e.g.,][]{conroy_etal06}. 
This is also consistent with the previous results that find that both 
radial and velocity biases of subhalos largely disappear if subhalos are selected
using their properties at the accretion epoch, unaffected by tidal stripping \citep{nag05,faltenbacher_etal06}.
Moreover, this explains why the bias is slightly higher in the inner region
(e.g., $\Delta_c=500$), where the tidal effects should be more
significant.  

The importance of how subhalos are selected is further highlighted in
Figure~\ref{fig:velpro}, which compares galaxy velocity dispersion
profiles ({\it points with errorbars}) to the velocity dispersion
profiles of dark matter particles, indicated by lines.  The figure
shows that velocity dispersions of galaxies are generally higher than
those of dark matter in the radial range $0.1<r/r_{200c}<1$.  The
velocity bias is slightly larger when galaxies are selected using their
current total mass, rather than their stellar mass.  The figure also
illustrates that both velocity and spatial biases are smaller at
higher redshifts, where the tidal stripping and disruption had not yet
had time to affect the masses of subhalos, and hence their spatial
distribution and overall velocity dispersion.

For high-redshift objects, the number of cluster galaxies $N_{\rm gal}$ is often too 
limited for reliable dynamical mass.  Using CSF simulations, we quantify how well one 
can estimate dark matter velocity dispersion using a limited number of galaxies.  
Specifically, we compute the velocity bias using 3, 5, 10 and 20 most massive galaxies 
(excluding the central galaxy) in our CSF halos. The results are summarized in Table~\ref{tab:vbias}.  
Our analysis shows that galaxy velocity dispersion tends to exhibit less bias for 
larger number of galaxies.  At  $z=0$, galaxy velocity dispersion is essentially unbiased 
(i.e., $b_v = 1$) within 1$\sigma$ for $N_{\rm gal} > 10$ at both 
$\Delta_c=200$ and $500$.  At higher redshifts, larger number of galaxies are required 
to achieve unbiased estimate.  Our analyses indicate that the galaxy velocity 
dispersions are unbiased within 2$\sigma$ for $N_{\rm gal} \geq 20$ at $z=0.6$ and $1$.  
A larger sample of simulated clusters are required to confirm the remaining bias at a 
level of 5\% at these redshifts.  Moreover, our results clearly show that the velocity bias 
is significant for $N_{\rm gal}<10$ galaxies.  This is because the 
more massive galaxies reside nearer to the inner region due to dynamical friction and 
hence have lower velocities. 

\begin{table}[t]
\caption{Velocity Bias}
\begin{center}
\begin{tabular}{l|ccc}
\hline
\hline
mean$\pm$1-$\sigma$ error&&$\Delta_c = 200$&\\
\cline{2-4}
&$z=0.0$&$z=0.6$&$z=1.0$\\
\hline
$b_{v,sub}^{\rm NR}$&$1.131\pm0.022$&$1.063\pm0.036$&$1.073\pm0.032$\\
$b_{v,sub}^{\rm CSF}$&$1.067\pm0.021$&$1.050\pm0.021$&$1.032\pm0.024$\\
$b_{v,gal}^{\rm CSF}$&$1.029\pm0.022$&$0.993\pm0.032$&$0.971\pm0.033$\\
$b_{v,gal}^{\rm CSF}\ (N_{\mathrm{gal}}=3)$&$0.796\pm0.070$&$0.792\pm0.055$&$0.772\pm0.065$\\
$b_{v,gal}^{\rm CSF}\ (N_{\mathrm{gal}}=5)$&$0.857\pm0.042$&$0.783\pm0.053$&$0.792\pm0.047$\\
$b_{v,gal}^{\rm CSF}\ (N_{\mathrm{gal}}=10)$&$0.964\pm0.035$&$0.894\pm0.027$&$0.856\pm0.047$\\
$b_{v,gal}^{\rm CSF}\ (N_{\mathrm{gal}}=20)$&$0.976\pm0.025$&$0.969\pm0.027$&$0.957\pm0.037$\\
\hline
\hline
&&$\Delta_c = 500$&\\
\cline{2-4}
&$z=0.0$&$z=0.6$&$z=1.0$\\
\hline
$b_{v,sub}^{\rm NR}$&$1.200\pm0.021$&$1.084\pm0.043$&$1.116\pm0.039$\\
$b_{v,sub}^{\rm CSF}$&$1.128\pm0.028$&$1.084\pm0.021$&$1.078\pm0.025$\\
$b_{v,gal}^{\rm CSF}$&$1.083\pm0.031$&$1.018\pm0.034$&$1.020\pm0.028$\\
$b_{v,gal}^{\rm CSF}\ (N_{\mathrm{gal}}=3)$&$0.811\pm0.077$&$0.756\pm0.066$&$0.816\pm0.067$\\
$b_{v,gal}^{\rm CSF}\ (N_{\mathrm{gal}}=5)$&$0.945\pm0.052$&$0.821\pm0.052$&$0.887\pm0.057$\\
$b_{v,gal}^{\rm CSF}\ (N_{\mathrm{gal}}=10)$&$1.013\pm0.043$&$0.928\pm0.037$&$0.958\pm0.039$\\
$b_{v,gal}^{\rm CSF}\ (N_{\mathrm{gal}}=20)$&$1.030\pm0.035$&$1.022\pm0.019$&$1.067\pm0.037$\\
\hline
\end{tabular}
\end{center}\label{tab:vbias}
\end{table}
 
\section{Summary and Discussion}
\label{sec:summary}

In this work we investigate the effect of baryon dissipation on the
scaling relation between total mass and velocity dispersion, 
and the velocity bias of galaxies in clusters. To this end,
we use a representative set of self-consistent cosmological cluster
simulations that follow the formation and evolution of galaxies by
taking into account effects of gas cooling, star formation, and
feedback.

We confirm that the virial scaling relation of dark matter exhibits a
remarkable degree of self-similarity with log-normal scatter of only
5\%.  The virial scaling relation in our non-radiative hydrodynamical
simulations is consistent with that of the dissipationless simulations
presented by \citet{evrard_etal08}.  This demonstrates that the virial
scaling relation has now been calibrated at a level of {\it a few percent} 
in the non-radiative regime.

We show that the largest source of systematic uncertainty in the
intrinsic virial scaling relation calibration lies in our
understanding of baryonic physics. Using numerical simulation that
include the physics of galaxy formation, we find that the baryon
dissipation increases dark matter velocity dispersion of of groups 
and clusters at the level of $\sim 4\%$ to $10\%$ due to adiabatic contraction of dark
matter in the central regions.  Our simulations indicate that this
effect is more pronounced in lower mass systems where gas cooling and
star formation are more efficient than massive systems.  Note,
however, that the effect of baryon dissipation in this work should be
taken as an upper limit, because the current simulations suffer from
the overcooling problem \citep[e.g.,][]{borgani_kravtsov09}. Despite
this problem, we find that the scatter in the virial scaling relation
is very tight (only $\sim 5\%$), independent of mass and redshift, and
insensitive to details to input cluster physics.

In practice, the dark matter virial scaling relation is not direclty
accessible to observations.  Instead, observers rely on galaxies as
tracers of the underlying distribution and dynamics of matter.  Thus,
the bias in the velocity dispersions of galaxies with respect to that
of dark matter is one of the main sources of systematic uncertainty in
the mass estimates of galaxy clusters.  Previous works based on
$N$-body simulations have shown that dark matter ``subhalos'' in
simulations exhibits a positive velocity bias, because slow subhalos
are much less common, due to physical disruption by gravitational
tides early in the merging history
\citep[e.g.,][]{col_etal00,diemand_etal04,gao_etal04}.

In this work, we show that the galaxy velocity bias, in fact, depends
on how galaxies (``subhalos'') are selected in simulations.  Using the
self-consistent cluster simulations, we find that galaxies selected
based on their stellar mass exhibit no velocity bias. Moreover, we find 
that the observational estimates of galaxy velocity dispersion with more 
than twenty most massive galaxies in groups and clusters is an unbiased 
tracer of the dynamics of underlying matter.

\acknowledgments
EL and AK are supported by the NSF grant
AST- 0708154, by NASA grant NAG5-13274, and by Kavli Institute for
Cosmological Physics at the University of Chicago through grant NSF
PHY-0551142 and an endowment from the Kavli Foundation. This work made
extensive use of the NASA Astrophysics Data System and arXiv.org
preprint server.

\bibliography{ms.bib}

\end{document}